\documentclass{article}
\usepackage[british]{babel}
\usepackage[breaklinks]{hyperref}

\usepackage{amsmath}
\usepackage{amssymb}

\newcommand{\lie}{\mathcal{L}}
\newcommand{\bmu}{\boldsymbol{\mu}}
\newcommand{\bsi}{\boldsymbol{\sigma}}

\begin{document} 

\pagenumbering{arabic}
\title{The homogeneous and isotropic Weyssenhoff fluid}
\author{Christian G. B\"ohmer\footnote{e-mail: {\tt boehmer@hep.itp.tuwien.ac.at}}\\
        \footnotemark[1]~ASGBG/CIU, Department of 
        Mathematics, Apartado Postal C-600\\
        University of Zacatecas (UAZ), Zacatecas, Zac 98060, Mexico
        \and
        Piotr Bronowski\footnote{e-mail: {\tt pbronowski@poczta.fm}}\\
        \footnotemark[2]~Institute for Theoretical Physics, 
        University of Lodz,\\ Pomorska 149/153, PL-90-236 Lodz, Poland.}
\date{}
\maketitle

\begin{abstract}
We consider a Weyssenhoff fluid assuming that 
the spacetime is homogeneous and isotropic, therefore being relevant
for cosmological considerations of gravity theories with torsion. 
In this paper, it is explicitely shown that the Weyssenhoff 
fluids obeying the Frenkel condition or the Papapetrou-Corinaldesi 
condition are incompatible with the cosmological principle, which 
restricts the torsion tensor to have only a vector and an axial 
vector component. Moreover it turns out that the Weyssenhoff fluid 
obeying the Tulczyjew condition is also incompatible
with the cosmological principle. Based on this result we propose to
reconsider a number of previous works that analysed cosmological solutions 
of Einstein-Cartan theory, since their spin fluids usually did not obey 
the cosmological principle.
\end{abstract}
\mbox{} 
\newpage
\section{Introduction}

Cosmology, over the last years, has become a very active field of
research containing many open questions that require further
investigation. Hawking, Penrose and others have shown that under
fairly general assumptions solutions of Einstein's field equations
evolve singularities. This, from a conceptual point of view, is
rather unsatisfactory, we refer the reader to~\cite{Hawking:1973}.

When cosmological models with torsion were first studied, it was 
hoped that the inclusion of torsion would help to avoid these
singularities. Unfortunately this could only be achieved assuming
quite unrealistic matter models, see e.g.~\cite{Hehl:1976kj}.
It turns out, however, that most of these cosmological models
with torsion did not satisfy the cosmological principle, sometimes
also known as the Copernican principle, that strongly restricts the 
metric and the torsion tensor.
 
The Copernican principle states that the universe is spatially
homogeneous and isotropic on very large scales. This principle
takes the following mathematically precise form. The 
four-dimensional $4d$ spacetime manifold $(\mathcal{M},g)$ is 
foliated by $3d$ constant time spacelike hypersurfaces, which are 
the orbits of a Lie group $G$ acting on $\mathcal{M}$, with isometry 
group $SO(3)$. Following the Copernican principle~\cite{Tsamparlis:1979}, 
we assume all fields to be invariant under the action of $G$
\begin{align}
      \lie_{\xi} g_{\mu\nu}=0\,,\qquad
      \lie_{\xi} T^{\lambda}{}_{\mu\nu}=0\,,
      \label{eq:cos1}
\end{align}
where $\xi$ are the (six) Killing vectors generating the spacetime
isometries. The metric tensor is denoted by $g_{\mu\nu}$ ,
$T^{\lambda}{}_{\mu\nu}$ denotes the torsion tensor and Greek
indices label the holonomic components. For the rest of the paper
only anholonomic components of tensors are used, labelled by Latin
indices.

Kopczy\'nski initiated the investigation of cosmological models
with torsion in Ref.~\cite{Kopczynski:1972} and Ref.~\cite{Kopczynski:1973}, 
who assumed a Weyssenhoff fluid to be the source of both curvature 
and torsion. In~\cite{Kopczynski:1972} a non-singular universe
with torsion was constructed and in~\cite{Kopczynski:1973} an
anisotropic model of the universe with torsion was analysed.
The cosmological principle in the above strict sense~(\ref{eq:cos1})
was first developed in Einstein-Cartan theory by Tsamparlis 
in~\cite{Tsamparlis:1979}, where it was also suggested to reconsider 
the results in~\cite{Kopczynski:1972,Kopczynski:1973},
since the Weyssenhoff fluid turns out to be incompatible with the
cosmological principle (see also~\cite{Obukhov:1987yu}). 
The spin tensor used by~\cite{Kopczynski:1973} in a cosmological
context had just one non-vanishing component $S_{23}=K$, where
$K$ was assumed to be a function of the time variable $K=K(t)$. 
Such a spin tensor, as we will show below, is not compatible with 
the cosmological principle, a fact that was noted by Kopczy\'nski. 
It should also be pointed out that if we require only the metric
of Friedman-Robertson-Walker (FRW) type and put no restrictions on 
the torsion tensor, then the Weyssenhoff fluid can consistently be 
used as a source for curvature and torsion (see the energy-momentum 
tensor Eq.~(5.2) and (5.3) in Ref.~\cite{Obukhov:1987yu}). However, 
by doing so, one must drop the second condition of~(\ref{eq:cos1}), 
$\lie_{\xi} T^{\lambda}{}_{\mu\nu}=0$ and use a weaker notion
of the cosmological principle. For a recent example where a 
so-called cosmological model with macroscopic spin fluid was analysed 
see Ref.~\cite{Szydlowski:2003nv}.

Applying the restrictions~(\ref{eq:cos1}) to $g_{\mu\nu}$ yields the 
FRW type metric
\begin{align}
      ds^2=dt^2 - \Bigl(\frac{a(t)}{1-\frac{k}{4}r^2}\Bigr)^2 (dx^2+dy^2+dz^2)
      \label{eq:cos2}
\end{align}
where $r^2=x^2+y^2+z^2$ and where the 3-space is spherical for $k=1$,
flat for $k=0$ and hyperbolic for $k=-1$.
If we impose the restrictions~(\ref{eq:cos1}) on the torsion 
tensor~\cite{Tsamparlis:1979}, its allowed components are
\begin{align}
      T_{xxt} = T_{yyt} = T_{zzt} = h(t)\,,
      \nonumber \\
      T_{xyz} = T_{zxy} = T_{yzx} = f(t)\,,
      \label{eq:cos5}
\end{align}
where we follow the notation of~\cite{Goenner:1984}.
Hence, the cosmological principle allows a vector torsion
component $T_t=T^a{}_{at}=3h$, along the world lines, and
an axial vector component $T_{xyz}=f\epsilon_{xyz}$, within the
hypersurfaces of constant time. Such a totally skew-symmetric
torsion tensor in cosmology was considered earlier
in~\cite{Minkowski:1986kv}, where $h=0$ was assumed. The geometry
parameter $k$ was redefined to include the remaining torsion by
$\bar{k}=k-f^2a^2/2$. Also with $h=0$, the cosmological inflation  
could be explained by torsion in Ref.~\cite{Boehmer:2005sw}, 
using a rough model.

\section{The cosmological field equations}

The spin-connection 1-form $\tilde{\omega}^i{}_j$ in theories with 
torsion can be split into a torsion free part (the usual spin-connection 
1-form $\omega^i{}_j$ related to the Christoffel symbol $\Gamma_{ij}^k$) 
and a contortion 1-form part $K^i{}_j$, that takes 
the torsion of spacetime into account
\begin{align}
      \tilde{\omega}^i{}_j = \omega^i{}_j + K^i{}_j\,,
      \label{eq:cos6}
\end{align}
where the torsion tensor and the contortion tensor are related
by the following algebraic relation
\begin{align}
      T^i &= De^i = de^i + \tilde{\omega}^i{}_j e^j =
      K^i{}_j \wedge e^j \,,
      \label{eq:cos6a}
\end{align}
where we used that $\omega^i{}_j$ is a torsion-free connection.
The latter relation between torsion and contortion also implies that 
their vector and axial vector component are simply related by
\begin{align}
      T_{[ijk]} = K_{[ijk]}\,, \qquad T_{ij}{}^j = \frac{1}{2} K_{ji}{}^j \,.
      \label{eq:cos6b}
\end{align}
Since the metric and the contortion (or torsion) components that
are compatible with the cosmological constant are fixed, one
can compute any geometrical quantity of interest.

Metric~(\ref{eq:cos2}) gives rise to the following basis 1-forms
\begin{align}
      e^t = dt\,, e^{x,y,z} = \frac{a(t)}{1-\frac{k}{4}r^2} dx,y,z\,,
\end{align}
which together with the non-vanishing torsion components~(\ref{eq:cos6})
yields the following non-vanishing connection 1-forms 
\begin{align}
      \omega^t{}_x &= \frac{\dot{a}}{a} e^x + h e^x \,, 
      \nonumber \\
      \omega^t{}_y &= \frac{\dot{a}}{a} e^y + h e^y \,, 
      \nonumber \\
      \omega^t{}_z &= \frac{\dot{a}}{a} e^z + h e^z \,, 
      \nonumber \\
      \omega^x{}_y &= \frac{ky}{2a}e^x - \frac{kx}{2a}e^y - \frac{f}{2}e^z \,,
      \nonumber \\
      \omega^x{}_z &= \frac{kz}{2a}e^x - \frac{kx}{2a}e^z + \frac{f}{2}e^y \,,
      \nonumber \\ 
      \omega^y{}_z &= \frac{kz}{2a}e^y - \frac{ky}{2a}e^z - \frac{f}{2}e^x \,,
      \label{eq:cos9a}
\end{align}
that are computed from ~(\ref{eq:cos6}) $T^i = de^i + \tilde{\omega}^i{}_j e^j$, 
where the torsion two form can be obtained from~(\ref{eq:cos6}) via 
$T^i = (1/2) T^i{}_{jk}e^j \wedge e^k$. The field equations of Einstein-Cartan 
theory~\cite{Hehl:1976kj} are obtained by varying the usual Einstein-Hilbert 
action with respect to the vielbein and the spin-connection as independent
variables
\begin{align}
      R^i{}_j - \frac{1}{2}R \delta^i_j = 8\pi\, \Sigma^i{}_j \,,
      \nonumber \\
      T^i{}_{jk}-\delta^i_j T^l{}_{lk}-\delta^i_k T^l{}_{jl} = 8\pi\, s^i{}_{jk} \,.
      \label{eq:cos9}
\end{align}
$\Sigma^i{}_j$ is the canonical energy-momentum tensor and
$s^i{}_{jk}$ is the tensor of spin.

\section{The Weyssenhoff fluid}

The ideal Weyssenhoff fluid~\cite{Weyssenhoff:1947} is a generalisation 
of the ideal fluid to take into account the properties of spin and torsion 
in spacetime. Its canonical energy-momentum tensor is given by
\begin{align}
      \Sigma_{ij}=p_{i}u_{j}+P (u_i u_j - g_{ij})\,,
      \nonumber \\
      \qquad p_i=\rho u_i - u^l \nabla_k (u^k S_{li})\,, 
      \label{eq:wey1} \\
      s^{i}{}_{jk}=u^{i}S_{jk}\,,
      \label{eq:wey2} 
\end{align}
where $p_i$ is the momentum density of the fluid and $u_i$
is the fluid's velocity. By $\rho$ and $P$ we denoted the 
energy density and the pressure of the fluid, respectively.
The intrinsic angular momentum tensor $S_{ij}$ satisfies      
\begin{align}
      S_{ij} = -S_{ji} \,.
      \label{eq:wey3}
\end{align}
The spin tensor $S^{ij}$ can be decomposed into two 3-vectors
\begin{align}
      \bmu := (S^{01},S^{02},S^{03}) \,, 
      \label{eq:wey3a}
\end{align}
that in case we assume the Frenkel condition~\cite{Frenkel:1926}, vanishes in the 
rest-frame. The second vector
\begin{align}
      \bsi := (S^{23},S^{31},S^{12}) \,,
      \label{eq:wey3b}
\end{align}
in the rest-frame can be regarded as the spin density.

Integrability of the particles' equations of motion requires one
more condition that the spin tensor has to satisfy
\begin{align}
      \zeta^{i}S_{ji}=0\,,
      \label{eq:wey4}
\end{align}
where the vector $\zeta^i$ is usually taken to be the velocity
vector of the fluid $u^i$, following Frenkel~\cite{Frenkel:1926}.
It is also possible to choose the momentum density according to
Tulczyjew~\cite{Tulczyjew:1959}. Another frequently used condition
was put forward by Papapetrou and Corinaldesi~\cite{Corinaldesi:1951pb}
who assumed the condition
\begin{align}
      S^{t i} = 0 \,,
      \label{eq:wey4b}
\end{align}
where with $t$ stands for the time component of the spin tensor.
In the following sections we investigate whether a Weyssenhoff
fluid obeying one of the three presented integrability conditions
is compatible with the cosmological principle.

\section{The Frenkel condition}

If we assume the Frenkel condition~\cite{Frenkel:1926} $\zeta^i = u^i$ 
then the spin contribution of the energy-momentum tensor can be 
rewritten to given
\begin{align}
      u^l \nabla_k (u^k S_{li}) = u^l S_{li} \nabla_k u^k + u^l u^k \nabla_k S_{li}
      \nonumber \\
      = u^l u^k \nabla_k S_{li} = -a^l S_{li} \,.
      \label{eq:wey4a}
\end{align}
In the third and fourth step the Frenkel was necessary for
the modifications, and we introduced the acceleration of the fluid 
$a^j$, defined by $a^j=(u^k \nabla_k)u^j$. Hence equations~(\ref{eq:wey1})
and~(\ref{eq:wey2}) taking the Frenkel condition into account yield
\begin{align}
      \Sigma_{ij}=\rho u_{i}u_{j}+P (u_i u_j - g_{ij}) + a^l S_{li} u_j \,,
      \nonumber \\
      s^{i}{}_{jk}=u^{i}S_{jk}\,, \qquad u^{i}S_{ik} = 0 \,.
      \label{eq:wey6} 
\end{align}
This implies that for vanishing acceleration $a^j$ of the fluid one 
is back at Einstein gravity~\cite{Griffiths:1982}. The interpretation of the
contribution of the spin angular momentum tensor in~(\ref{eq:wey1})
in terms of the acceleration strongly depends on the Frenkel condition.

The totally skew-symmetric part of the torsion tensor~(\ref{eq:cos5}) 
is allowed by the cosmological principle. Since the four velocity $u^i$ 
enters the definition of the tensor of spin~(\ref{eq:wey2}), a Weyssenhoff 
like fluid cannot be the source of the totally skew-symmetric torsion 
component~(\ref{eq:cos5}). On the other hand, it is the Frenkel 
condition~(\ref{eq:wey4}) with $\zeta^i=u^i$ which does not allow 
the Weyssenhoff fluid  to be the source of the trace components
(\ref{eq:cos5}) of the torsion tensor. More explicitely, multiplying 
the torsion field equation~(\ref{eq:cos9}b) by $\delta^j_i$ leads to
\begin{align}
      -2 T^l{}_{lk} = 8\pi s^l{}_{lk} = 8\pi u^{l}S_{lk} = 0\,,
      \label{eq:wey7}
\end{align}
where for the last steps~(\ref{eq:wey2}) and the Frenkel condition 
were taken into account. Therefore we have explicitely shown that
the Weyssenhoff fluid obeying the Frenkel condition is incompatible 
with the cosmological principle put forward in Ref.~\cite{Tsamparlis:1979}.

\section{The Papapetrou-Corinaldesi condition}

Assuming the Papapetrou-Corinaldesi~\cite{Corinaldesi:1951pb} condition 
$S^{t j}$ has the following consequences for the torsion tensor 
implied by the spin fluid. As before, since the fluid's four velocity 
enters the definition of the spin tensor, the totally skew-symmetric 
torsion has to vanish. Secondly, the traced torsion field 
equation~(\ref{eq:cos9}b) yields
\begin{align}
      -2 T^l{}_{lk} = 8\pi s^l{}_{lk} = 8\pi u^{l}S_{lk} 
      = 8\pi \delta^{l}_{t} S_{lk} = 8\pi S_{t k} = 0\,,
      \label{eq:wey7a}
\end{align}
where the vanishing of the trace part of the torsion tensor
is identically the condition of Papapetrou-Corinaldesi~(\ref{eq:wey4b}).

Therefore we again conclude that also the Weyssenhoff fluid obeying
the Papapetrou-Corinaldesi condition is incompatible with the 
cosmological principle (since we have $u^i=\delta_0^i$ the 
Papapetrou-Corinaldesi and the Frenkel condition in fact take 
the same form).

\section{The Tulczyjew condition}

According to our information, it has not been analysed so far if
the Weyssenhoff fluid obeying the Tulczyjew condition~\cite{Tulczyjew:1959}
is compatible with the cosmological principle. As in the
previous section, the fluid cannot be a source of the totally 
skew-symmetric component of the torsion tensor, because the four
velocity $u^i$ of the fluid is present in Eq.~(\ref{eq:wey2}). 
However, in this case~(\ref{eq:wey4}) does not vanish identically
on general ground and we arrive at
\begin{align}
      -2 T^l{}_{lk} = 8\pi s^l{}_{lk} = 8\pi u^{l}S_{lk} \,,
      \label{eq:wey8}
\end{align}
where the last term on the right hand side need not to vanish. 
The Tulczyjew condition, Eq.~(\ref{eq:wey4}) with $\zeta^i = p^i$
explicitely written out leads to
\begin{align}
      S_{ij} p^j = S_{ij} \bigl(\rho u^j - u_l \nabla_k (u^k S^{lj})
      \bigr) = 0 \,,
      \label{eq:wey9}
\end{align}
which can be used to express the last term on~(\ref{eq:wey8}) by
\begin{align}
      \rho S_{ij} u^j = S_{ij} u_l \nabla_k (u^k S^{lj}) \,.
      \label{eq:wey10}
\end{align}
The fluid's four velocity simply reads $u^j=\delta^j_0$, and
equation~(\ref{eq:wey10}) by taking~(\ref{eq:cos9a}) into account 
yields the following form of the Tulczyjew condition that the
spin fluid has to satisfy
\begin{align}
      \rho S_{i0} = S_{ij} \Bigl( \dot{S}_{0j} + \Gamma_0 S_{0j} \Bigr)\,,
      \label{eq:wey11} 
\end{align}
where the dot means differentiation with respect to time $t$.
In contrast to the two previous cases we find that this condition
does not imply the vanishing of the resulting trace of the torsion 
tensor. For the trace of the Christoffel symbol we find 
$\Gamma_0=\Gamma^k_{k0}=3(H+h)$ where by $H$ we denoted the Hubble
parameter defined by $H=\dot{a}/a$. Equations~(\ref{eq:wey11})
provide us with for conditions ($i=0,1,2,3$) which we will analyse
in more detail. For $i=0$ the left-hand side of~(\ref{eq:wey11})
vanishes and one is left with
\begin{align}
      0 &= S_{0j} \Bigl( \dot{S}_{0j} + \Gamma_0 S_{0j} \Bigr) 
      \nonumber \\
        &=-S^{0j} \Bigl( \dot{S}_{0j} + \Gamma_0 S_{0j} \Bigr)\,,
      \label{eq:wey11a}
\end{align}
which can be written in an equivalent form by using the
introduced vectors $\bmu$ and $\bsi$ in~(\ref{eq:wey3a})
and~(\ref{eq:wey3b}) and leads to
\begin{align}
      0 = \bigl( \dot{\bmu} + \Gamma_0\, \bmu \bigr) \cdot \bmu \,,
      \label{eq:wey12}
\end{align}
where by $\cdot$ we mean the usual inner product of vectors.
From this we conclude that the vectors $\bmu$ and $(\dot{\bmu} + \Gamma_0 \bmu)$
are orthogonal to each other.
For the remaining values $i=1,2,3$ the condition~(\ref{eq:wey11})
takes the following form in terms of the three vector
\begin{align}
      \bmu = \frac{1}{\rho}\bigl( \dot{\bmu} + \Gamma_0\, \bmu \bigr) \times \bsi \,.
      \label{eq:wey13}
\end{align}
Here we now see that~(\ref{eq:wey12}) is not an independent
equation since from the latter we derive that $\bmu\perp(\dot{\bmu}+\Gamma_0\bmu)$
and moreover that $\bmu\perp\bsi$.
Therefore we find the following non-vanishing components of
the induced torsion tensor via the field equations~(\ref{eq:wey8})
\begin{align}
      T^l{}_{l0} &= 0 \,,
      \nonumber \\
      T^l{}_{li} &= 4\pi \bmu \,,
      \label{eq:wey14}
\end{align}
where $\bmu$ is given by equation~(\ref{eq:wey13}). Note that the
index $i$ only takes the values $1,2,3$ and that we furthermore
suppressed the explicit index for the vector $\bmu$. We can now
try to continue the construction of a spin fluid that is compatible with
cosmological principle, namely the Weyssenhoff fluid obeying the
Tulczyjew condition. In principle we have two possibilities: ({\bf a})
we choose the vector $\bmu$ of the spin tensor so that it satisfies
the condition~(\ref{eq:wey12}) and we choose the spin density vector 
$\bsi$ so that equation~(\ref{eq:wey13}) is satisfied. 
On the other hand, ({\bf b}) let us prescribe the spin 
density vector $\bsi$. Then, in order to get an allowed $\bmu$,
one has to solve the vector differential equation~(\ref{eq:wey13}),
so that each solution satisfies~(\ref{eq:wey12}). However, one must
be careful with the above result. The cosmological principle allows 
a vector torsion component, but only along the world lines, $T^l{}_{lt}\neq 0$.
The other components are excluded. Therefore, also the Weyssenhoff fluid
obeying the Tulczyjew condition is incompatible with the cosmological
principle, since~(\ref{eq:wey14}a) vanishes identically.

\section{Conclusions and outlook}

The restrictions that follow from assuming homogeneity and isotropy on
the very large scales of the universe (the cosmological principle) allow 
one metric component and two torsion components; a vector and an axial vector 
component of the torsion tensor. We showed that the Weyssenhoff fluids obeying 
either the Frenkel or the Papapetrou-Corinaldesi condition are incompatible with 
this principle. Furthermore we analysed the Tulczyjew condition which, in 
principle, allows one to construct a non-trace-free torsion tensor. However, 
its time component, allowed by the cosmological principle, vanishes identically. 
Therefore it has been shown that no spin fluid obeying the common integrability 
conditions is compatible with the cosmological principle. This rather 
surprising result shows the necessity to reconsider the previous works
on cosmology with torsion, since none of these results can be regarded
as a truly cosmological model with torsion.

Furthermore it raises the question, whether an integrability condition exists 
that allows a spin fluid to have homogeneous and isotropic
torsion components. The construction of such a spin fluid, if possible,
could be the subject of further research. If it turned out that such a spin
fluid does not exist, this would have quite significant consequences
for the physical applicability of such models. The possible non-existence
would indicate that the Weyssenhoff fluid is not a very good model for
a macroscopic spin fluid. If, on the hand, such a cosmological spin fluid
can be constructed, it would be very interesting to study its properties.
For example, the consequences of a truly cosmological spin fluid on the
singularities, mentioned in the introduction, were worth a thorough
investigation. Moreover it would then be possible to reconsider some
of the previously suggested models in a real cosmological fashion.
Finally we would like to mention the possibility of applying the cosmological 
principle to the more general hyperfluid~\cite{Obukhov:1993pt,Obukhov:1996mg}. 
However, a axial vector component for the torsion tensor cannot be obtained 
from the hyperfluid, since a generalised form of equation~(\ref{eq:wey2}) 
essentially enters the tensor of spin.

\section*{Acknowledgements}
This work is partially supported by the Polish Ministry of Scientific Research 
and Information Technology under the grant \mbox{No. PBZ/MIN/008/P03/2003} 
and by the University of Lodz.
The work of CGB was supported by research grant BO 2530/1-1 of the
German Research Foundation (DFG).
\mbox{}\\
This work is part of the research project Nr. 01/04 
{\it Quantum Gravity, Cosmology and Categorification} 
of the Austrian Academy of Sciences (\"OAW) and the National
Academy of Sciences of Ukraine (NASU).

\end{document}